\documentclass{sigchi}

\toappear{\scriptsize Permission to make digital or hard copies of all or part of this work for personal or classroom use is granted without fee provided that copies are not made or distributed for profit or commercial advantage and that copies bear this notice and the full citation on the first page. Copyrights for components of this work owned by others than ACM must be honored. Abstracting with credit is permitted. To copy otherwise, or republish, to post on servers or to redistribute to lists, requires prior specific permission and/or a fee. Request permissions from permissions@acm.org. \\
	{\emph{CHI 2016}}, May 7--12, 2016, San Jose, California, USA. \\
	Copyright is held by the owner/author(s). Publication rights licensed to ACM. \\
	ACM ISBN 978-1-4503-3362-7/16/05\ ...\$15.00.\\
	DOI: \url{http://dx.doi.org/10.1145/2858036.2858353}}

\usepackage{balance}  %

\usepackage{times}

\usepackage{url}      %

\makeatletter
\def\url@leostyle{%
  \@ifundefined{selectfont}{\def\UrlFont{\sf}}{\def\UrlFont{\small\bf\ttfamily}}}
\makeatother
\def\pprw{8.5in}
\def\pprh{11in}

\usepackage{booktabs}
\usepackage{paralist}
\usepackage{fixltx2e}
\usepackage{subfigure}
\usepackage{listings}
\usepackage{float}
\usepackage{caption}
\usepackage{soul}
\usepackage[normalem]{ulem}

\usepackage{graphicx}
\usepackage{algorithm2e}
\usepackage[noend]{algorithmic}
\usepackage{amsfonts}
\usepackage{amsmath}
\usepackage{ifthen}
\usepackage[usenames]{xcolor, colortbl}
\usepackage{amssymb}
\usepackage{dblfloatfix}
\usepackage{bm}
\usepackage{microtype}

\usepackage{tweaklist}
\renewcommand{\descripthook}{\setlength{\topsep}{0.2\baselineskip}%
  \setlength{\itemsep}{0.1\baselineskip}}
\renewcommand{\itemhook}{\setlength{\topsep}{-0.5\baselineskip}%
  \setlength{\itemsep}{0.1\baselineskip}}

\definecolor{linkColor}{RGB}{6,125,233}
\usepackage[pdftex]{hyperref}
\hypersetup{
pdftitle={Airways: Optimization-Based Planning of Quadrotor Trajectories according to High-Level User Goals},
pdfauthor={Christoph Gebhardt, Benjamin Hepp, Tobias Nageli, Stefan Stesvic, Otmar Hilliges},
pdfkeywords={robotics; quadrotors; computational design; videography; CHI},
bookmarksnumbered,
pdfstartview={FitH},
colorlinks,
citecolor=black,
filecolor=black,
linkcolor=black,
urlcolor=linkColor,
breaklinks=true,
}

\newcommand{\hidden}[1]{}

\newboolean{showComments}
\setboolean{showComments}{true}
\definecolor{gold}{rgb}{0.80,.60,0}

\newcommand{\todo}[2]{
\ifthenelse{\boolean{showComments}}
{\framebox[\columnwidth][l]{\parbox[l]{3.25in}{\textcolor{red}{#1: #2}}}}
{}
}

\newboolean{showEdits}
\setboolean{showEdits}{false}
\definecolor{amber}{rgb}{1.0, 0.49, 0.0}
\definecolor{bittersweet}{rgb}{1.0, 0.44, 0.37}

\newcommand{\editsCR}[2]{%
\ifthenelse{\boolean{showEdits}}%
{\textcolor{bittersweet}{#1}}%
{#1}%
}

\newboolean{coloredDiff}
\setboolean{coloredDiff}{false}

\newcommand{\ccDiff}[1]{%
\ifthenelse{\boolean{coloredDiff}}%
{\textcolor{bittersweet}{#1}}%
{#1}%
}

\definecolor{LightCyan}{rgb}{0.88,1,1}
\definecolor{LightGreen}{rgb}{0.77,0.93,0.8}
\definecolor{LightOrange}{rgb}{0.95,.67,0.47}
\usepackage[first=0,last=9]{lcg}

\newcolumntype{g}{>{\columncolor{LightGreen}}l}
\newcolumntype{o}{>{\columncolor{LightOrange}}l}

\newcommand{\figref}[1]{Fig.~\ref{#1}}

\usepackage{letltxmacro}
\LetLtxMacro{\originaleqref}{\eqref}
\renewcommand{\eqref}{Eq.~\originaleqref}

\begin{document}

\sloppy
\title{Airways: Optimization-Based Planning of Quadrotor Trajectories according to High-Level User Goals}

\teaser{
	\centering{
		\includegraphics[width=\linewidth]{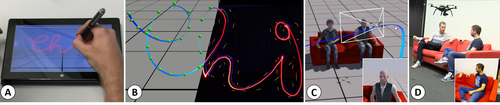}
		\caption{Interactive computational design of quadrotor trajectories: (A) user interface to specifiy keyframes and dynamics of quadrotor flight. (B) An optimization algorithm generates feasible trajectories and (C) a 3D preview allows the user to quickly iterate on them. (D) The final motion plan can be flown by real quadrotors. The tool enables the implementation of a number of compelling use cases such as (B) robotic light-painting, aerial racing and (D) aerial videography.}\label{fig:teaser}}
}

\numberofauthors{5}
\author{
	Christoph Gebhardt\footnotemark[2] , Benjamin Hepp\footnotemark[2] , Tobias N\"ageli, Stefan Stev\v{s}i\'{c}, Otmar Hilliges\\
	\affaddr{Departement of Computer Science, AIT Lab}\\
	\affaddr{ETH Z\"urich}\\
}
\maketitle

\begin{abstract}
In this paper we propose a computational design tool that allows end-users to create advanced quadrotor trajectories with a variety of application scenarios in mind. Our algorithm allows novice users to create quadrotor based use-cases without requiring deep knowledge in either quadrotor control or the underlying constraints of the target domain. To achieve this goal we propose an optimization-based method that generates feasible trajectories which can be flown in the real world. Furthermore, the method incorporates high-level human objectives into the planning of flight trajectories. An easy to use 3D design tool allows for quick specification and editing of trajectories as well as for intuitive exploration of the resulting solution space. We demonstrate the utility of our approach in several real-world application scenarios, including aerial-videography, robotic light-painting and drone racing. %
\end{abstract}

\footnotetext[2]{The first two authors contributed equally to this work.\label{note1}}


\subsection{ACM Classification Keywords}
I.2.9 Robotics: Autonomous vehicles; Operator interfaces; H.5.2 User Interfaces;

\section{INTRODUCTION}\label{sec:intro}
In recent years micro-aerial vehicles (MAVs), in particular Quadrotors, have seen a rapid increase in
popularity both in research and the consumer mainstream.
While the underlying mechatronics and control aspects are complex, the recent emergence of simple to
use hardware and easy programmable software platforms has opened the door to widespread adoption and
enthusiasts have embraced MAVs such as the AR.Drone or DJI Phantom in many compelling scenarios
including aerial photo- and videography. Furthermore, the HCI community has begun to explore these
drones in interactive systems such as sports assistants
\cite{Higuchi:2011:FSA,Mueller:2015,Nitta:2014:HAS} or display
\cite{Scheible2013c} of content.

Clearly there is a desire to use such platforms in a variety of application scenarios. Current SDKs
already give novices access to manual or waypoint based control of MAVs, shielding them from the
underlying complexities. %
However, this simplicity comes at the cost of flexibility. For instance, flying a smooth, spline-like
trajectory or aggressive flight maneuvers%
, for example to create an aerial light show (e.g.,
\cite{ArsElectronica2012,CannesFestival:2012:QuadShow}), is tedious or impossible with waypoint based
navigation. These limits exist because manufacturers place hard thresholds on the dynamics to ensure
flight stability for inexperienced pilots.

More importantly, state-of-the-art technologies offer only very limited support for users who want to
employ MAVs to reach a certain high-level goal. This is maybe best illustrated by the most successful
application area -- that of aerial videography. What a few years ago was limited to professional camera
crews, requiring cost-intensive equipment like a helicopter, can now, in principle, be done by end-users with a MAV
and an action camera. However, producing high-quality aerial footage is not an easy task -- it demands
attention to the creative aspects of videography such as frame composition and camera motion
(cf.~\cite{mascelli1998five}). In the case of airborne cameras, an operator needs to fly smoothly,
accurately and safely around a camera target. Furthermore, the target has to be framed properly
alongside further creative considerations. Thus this is a difficult task and typically requires at
least two experienced operators -- one pilot and a camera man
(cf.~\cite{Diaz:2015}). Our method tackles this problem by enabling a single novice user to  
fly challenging trajectories and still create aesthetically pleasing aerial footage.

\begin{figure*}[bth]
	\centering
	\includegraphics[width=1.0\linewidth]{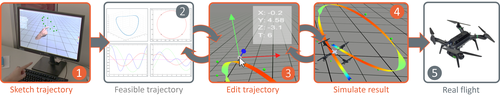}
	\caption{System workflow schematically. (1) User sketches keyframes. (2) An optimization method generates a feasible trajectory. (3+4) The user can quickly iterate over the trajectory and explore the solution space of feasible trajectories via a physics simulation or a rendered preview (see \figref{fig:sys_overview_camera}, D). (5) Final trajectory can be flown with a real MAV.}
	\label{fig:sys_overview}
\end{figure*}

\subsection{Overview \& Contribution}
Embracing the above challenges we propose a computational method that enables novice end-users to create quadrotor use-cases without requiring expertise in \textit{either} low-level quadrotor control \textit{or} specific knowledge in the target domain. The core contribution of our method is an optimization-based solution that generates \emph{feasible trajectories} for flying robots while taking \textit{high-level user goals} such as visually pleasing video shots, optimal racing trajectories or aesthetically pleasing MAV motion into consideration.
Furthermore, we develop an easy-to-use tool that allows for straightforward specification of flight trajectories and high-level constraints. Our approach guides the users in exploring the resulting design space via a 3D user interface and allows for quick iteration until finding a solution which fits best with the user's intentions. %

We demonstrate the flexibility of our approach in three real-world scenarios including aesthetically pleasing aerial-videos, robotic light-painting and drone racing.

\section{RELATED WORK}\label{sec:rw}
\subsection{MAVs in HCI}
With MAVs becoming consumerized the HCI community has begun to explore this design space. FollowMe
\cite{Naseer2013} is a MAV that follows a user and detects simple gestures via a depth camera, whereas
others have proposed using head motion for MAV control \cite{Higuchi:2012:FHH:2407707.2407719},
while \cite{Scheible2013c} propose a simple, remote controlled flying projection platform. Several setups have been proposed that turn such MAVs into flying, personal companions. For example, to
act as jogging partner \cite{Mueller:2015} or general purpose sports coach \cite{Higuchi:2011:FSA}, or
as an actuated and programmable piece of sports equipment~\cite{Nitta:2014:HAS}.

Commercially available  drones, targeted at the consumer market, shield the user form low-level flight aspects and provide simple manual control (e.g.\ using smartphones as controller) or waypoint based programmatic navigation as well as GPS based person
following. This dramatically lowers the entry barrier for novices but also limits the ceiling of
achievable robotic behavior. 
Our approach also aims for simplicity but gives more power to the users, enabling even novices to design and implement complex flight trajectories, concentrating on the high-level goals of the application domain.

\subsection{Video Stabilization \& Camera Path Planning}
Improving the visual quality of end-user produced content is a goal we share with post production video
stabilization. Inspired by early work which formulates the problem and discusses the aesthetics of
cinematography \cite{Gleicher:2008:RIC} a number of approaches employ computer vision methods to
estimate the original, jerky camera path. Based on this a new, smooth path is computed to
generate stabilized video \cite{Grundmann:2011,Liu:2013:BCP} and even time-lapse
footage~\cite{Kopf:2014:FHV} from the source material. 
Camera path planning has also been
studied extensively in the context of virtual environments using constraint based
\cite{christie2005virtual,CAV:CAV398} or probabilistic \cite{Li:2008:SG} methods. 
However, these methods are not limited by
real-world physics and hence can produce arbitrary camera trajectories and viewpoints. Our approach
differs from the above as we propose a \emph{forward} method that gives the user full control over the
creative aspect of camera planning while simultaneously optimizing for physical feasibility of the
flight path and cinematographic objectives. A 3D simulation lets the user explore the design space
before flying the actual trajectory and hence helps in understanding the trade-offs to
consider.

\subsection{Computational Design}
Sharing the goal of unlocking areas that previously required significant domain knowledge to novice
users, the HCI and graphics communities have proposed several methods that give novice users control
over aesthetic considerations while achieving functionality.
Recent examples include digitally designed gliders
\cite{Umetani2014} and kites \cite{Martin:2015:ODO} with optimized aerodynamic properties.
At the core of these approaches are sophisticated simulations or analytical models of the problem
domain that carefully balance accuracy and rapid responses to ensure interactivity while maintaining
guarantees (e.g., physical stability). We build on domain knowledge from the robotics and MAV
literature and propose an interactive design tool for complex MAV behavior usable by non-experts.  

\subsection{Robotic Behavior and Trajectory Generation}
Automating the design of robotic systems based on high-level functional specifications is a
long-standing goal in robotics, graphics and HCI. Focussing on robot behavior only, simple direct touch
and tangible UIs \cite{Yoshida2015}, and sketch based interfaces to program robotic systems
\cite{Liu:2011:RMS,Sakamoto2009} have been
proposed. Visual markers have been used to control robots explicitly, for example as kitchen aides
\cite{Sugiura2010}, or implicitly \cite{Kato:2012,Zhao:2009:MCP}, to schedule tasks for robots in-situ
which are then collected and executed asynchronously. 

Generating flight trajectories for MAVs is well-studied in robotics. In particular, the control aspects
of aggressive and acrobatic flight is an active area of research (e.g., \cite{Lupashin2012}).
Mellinger et al.'s work on generating minimum snap trajectories \cite{Mellinger2011a} is the most related to ours.                                  
While they specify a trajectory as a piecewise polynomial spline between keyframes, we discretize the trajectory
into small piecewise linear steps. The result is a more intuitive formulation of the optimization problem, making the incorporation of additional constraints and objectives much easier. We extend the approach in \cite{Mellinger2011a} by optimizing trajectories for flyability \textit{and} for high-level human objectives. We place
the users in the loop and provide easy-to-use tools to design quadrotor trajectories according to high-level objectives. 

Joubert et al.  \cite{Joubert2015} share a similar goal in proposing a design tool that allows novice users to
specify a camera trajectory, simulate the result, and execute the motion plan. In contrast to our work,
their method does not automate feasibility checking but delegates the correction of violations to
the user. Furthermore, our method allows to treat keyframes as soft instead of hard constraints,
allowing to trade off feasibility against keyframe matching. We also incorporate a larger number of high-level user constraints, such as additional cinematographic goals and collision-free trajectories, into the algorithm -- requiring a different formulation of the optimization problem. Finally, we demonstrate the gain in generality in our approach in the additional use cases of light writing and aerial racing.

\section{SYSTEM OVERVIEW}\label{sec:sys_overview}
We propose an end-to-end system that allows users to generate motion plans for quadrotors that are
'flyable' and adhere to high-level human-specified objectives for a variety of application
scenarios. \figref{fig:sys_overview} illustrates the design process. 

Using for example a LeapMotion
controller the user specifies keyframes, each consisting of a position and a time (1). The optimization
algorithm generates an initial 'flyable' trajectory from these inputs, i.e., one that lies within the
physical capabilities of the underlying quadrotor hardware (2). 
The method aims to find a solution that
goes through all specified keyframes,
however the user
may now adjust both a keyframe's position and timing as well as other parameters such as the overall flight
time, the optimization's objective (e.g., minimization of velocity) or the extend to which the generated
trajectory should follow the optimization's objective versus the position of the specified
keyframes (3). This can result in trajectories that do not directly meet the user inputs but are the best
trade-off between the potentially conflicting use-case specific constraints. 
A built-in physics simulation allows the user to virtually fly the
quadrotor and thus provides a better understanding of the expected real-world behavior and enables rapid iteration of trajectories (4). 
This tool already enables the design of various flight-maneuvers for example designing an aerial race-course or a
light-show (please see video for an illustration of the design process and results).

\begin{figure}[h]
	\centering
	\includegraphics[width=1.0\columnwidth]{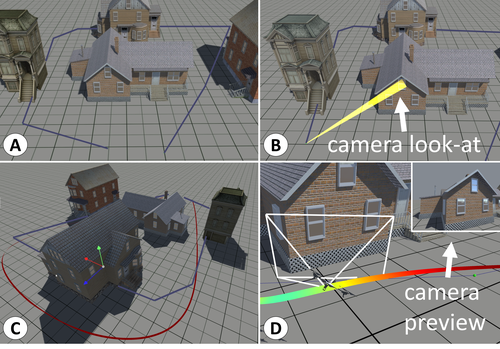}
	\caption{Planning of aerial video shots. (A) User specifies sparse keyframe positions connected by straight lines (purple). (B) For each keyframe the user also specifies a desired camera target (yellow). (C) We generate a smooth and collision free motion plan alongside gimbal control inputs (dark red). (D) Virtual preview allows for rapid prototyping showing the current camera frustum and a camera preview.}
	\label{fig:sys_overview_camera}
\end{figure}

However, the goal of our work is to enable more complex end-user scenarios as well. To this end we have
extended our method to also integrate high-level aesthetic constraints that are not necessarily
directly associated with the basics of quadrotor control. \figref{fig:sys_overview_camera} illustrates
how our tool can be used to plan aerial videography shots. In this case, the user designs an initial
camera trajectory around one or several targets. In addition to the keyframes the user specifies
targets which shall be captured by the on-board camera (\figref{fig:sys_overview_camera}, B). Our
algorithm generates both a quadrotor trajectory and a gimbal trajectory within the physical bounds. To
acquire visually pleasing footage our method incorporates cinematographic constraints such as smooth camera and target
motion, smooth transitions between multiple targets and reduction of perspective
distortions. Furthermore, the algorithm takes obstacle information into account and automatically
routes the trajectory through free-space (see \figref{fig:sys_overview_camera}, C). It would also be
straightforward to integrate other constraints such as limits of the coverage of a tracking system or government flight regulations.

In order to better understand the implications of the camera planning our tool allows the user to
virtually fly the shot by dragging the virtual quadrotor along the trajectory
(\figref{fig:sys_overview_camera}, D). For each point in time the tool renders the scene as it would be
captured in reality. The user can then edit the plan and iterate over different alternatives. 
Once satisfied the generated trajectories for quadrotor and gimbal can be deployed as a reference
to be followed by a real quadrotor.

\section{METHOD}\label{sec:method}
So far we have discussed the proposed design tool at a high-level and focused on how the user
accomplishes certain tasks.

We now introduce the underlying method we use to generate trajectories. To be able to reason about flight plans computationally, a model of the quadrotor and its dynamics are needed. This is a complex and challenging topic and we refer the reader to the Appendix for the full non-linear model that is needed to control the position and dynamics of the robot during flight (please see also \cite{Mahony2012}). The full model directly relates the inputs of a quadrotor to its dynamics -- this however makes trajectory generation a challenging problem and integrating such a highly non-linear model into an optimization scheme is complicated, incurs high computational cost and negates convergence guarantees \cite{Mellinger2011a}. However, for most application scenarios considered here a full non-linear treatment is not necessary as demonstrated by our results. In particular if the goal is to generate trajectories only (i.e., position and velocities) rather than the full control inputs as in \cite{Mellinger2011a}. Therefore, we present a linear
approximative model of the quadrotor and detail the optimization-based algorithm based on it.

\begin{figure}[b]
	\centering
	\includegraphics[width=0.6\linewidth]{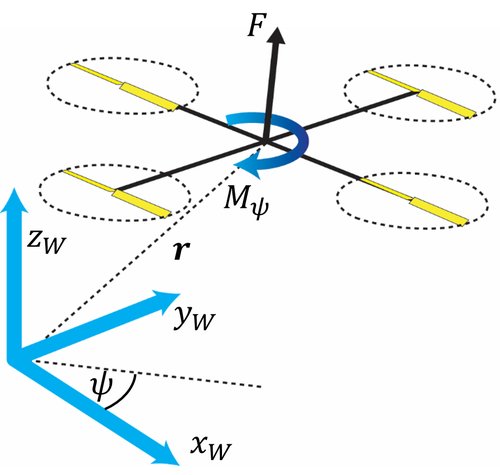}
		\caption{Our approximated quadrotor model with position $\mathbf{r}$, yaw angle $\psi$), 
		world frame ($\mathbf{x}_{W}$, $\mathbf{y}_{W}$, $\mathbf{z}_{W}$), the moment acting on the quadrotor along
the world frame z-axis $M_{yaw}$ and the force $F$ acting on the center of mass of the quadrotor.}
	\label{fig:quadrotor}
\end{figure}

\subsection{Approximate Quadrotor Model for Trajectory Generation}
\label{sec:sub:approximate_quad_model}
When generating a trajectory we want to ensure that it can be followed by a quadrotor,
i.e.\ a flight plan where each specified position and velocity can be reached within the time limits without exceeding the limits of the qudrotors	 inputs.

Therefore, we chose to approximate the quadrotor as a rigid body, described by its moment of inertia only along the world frame z-axis (i.e.\ we ignore pitch and roll of the quadrotor):
\begin{align}
\label{eq:quad_dynamical_system_continuous}
m\ddot{\mathbf{r}} = \mathbf{F} + m \mathbf{g} & \in \mathbb{R}^3 \\
I_{\psi} \ddot{\psi} = M_{\psi} & \in \mathbb{R} , \nonumber
\end{align}
where $\mathbf{r}$ describes the center of mass of the quadrotor,
$\psi$ is the yaw angle of the quadrotor, $m$ is the mass of the quadrotor,
$I_{\psi}$ is the moment of inertia about the body-frame z-axis,
$\mathbf{u}_{r}$ is the the force acting on the quadrotor
and $M_{\psi}$ is the torque along the z-axis.
This approximation allows to generate trajectories in the \textit{flat output} space of the
full quadrotor model (see Appendix for more details). 

In addition to the equations of motion we introduce bounds on the maximum
achievable force and torque:
\begin{equation}
\mathbf{u}_{\mathit{min}} \leq \mathbf{u} \leq \mathbf{u}_{\mathit{max}} \in \mathbb{R}^4  ,
\label{eq:input_bounds}
\end{equation}
where $\mathbf{u} = [\mathbf{F}, M_{\psi}]^{T}$ is the input of the system.

With this model it is not possible to exploit the full dynamic agility of a quadrotor.
As an example, consider the situation of accelerating straight upwards  by rotating all motors at maximum speed. To now also rotate around the body-frame z-axis we would have to lower the speed of motors $2$ and $4$, reducing the total thrust of the quadrotor. Currently we do not incorporate this coupling between the translational
and rotational dynamics into the bounds \eqref{eq:input_bounds} of our approximate
linear model \eqref{eq:quad_dynamical_system_continuous}. 
Therefore, to still ensure that a quadrotor can follow trajectories generated on base of this approximation conservative bounds are required.
Nonetheless, our results and applications demonstrate that these bounds still allow the quadrotor's agility to be sufficiently rich for many use cases. 
We refer the interested reader to the Appendix for details on how to choose these bounds.

For trajectory generation we rewrite the approximate model as a first-order dynamical system and discretize it in time with a time-step $\Delta t$
assuming a zero-order hold strategy, i.e.\ keeping the inputs constant in between stages:
\begin{equation}
\label{eq:quad_dynamical_system_discrete}
\mathbf{x}_{i+1} = A_d \mathbf{x}_{i} + B_d \mathbf{u}_{i} + c_{d} ,
\end{equation}
where $\mathbf{x}_{i} = [\mathbf{r}, \psi, \dot{\mathbf{r}}, \dot{\psi}]^{T} \in \mathbb{R}^{4}$
is the state and $\mathbf{u}_{i}$ is the input of the system at time $i \Delta t$. The matrix
$A_d \in \mathbb{R}^{8x8}$ propagates
the state $\mathbf{x}$ forward by one time-step, the matrix
$B_d \in \mathbb{R}^{8x4}$ describes the effect of the
input $\mathbf{u}$ on the state and the vector
$c_{d} \in \mathbb{R}^{8}$ that of gravity after one time-step.

\subsection{Trajectory Generation}
With this approximate quadrotor model in place we can now discuss the optimization scheme to generate
trajectories. The user specifies $M$ keyframes
describing a desired position $k_{j}$ 
at a specific time-point $\eta(j) \Delta t$, where
$\eta: \mathbb{N} \rightarrow \mathbb{N}$ maps the index of the keyframe to the corresponding time-point.
In the case of mouse-based user input we assume constant
time between consecutive positions. To compute a feasible trajectory  over the whole time horizon
$[0, t_{f}]$ we discretize time with a time-step $\Delta t$ into $N$ stages.
The variables we optimize for
are the quadrotor state $x_{i}$ and the inputs $u_{i}$ of the system
\eqref{eq:quad_dynamical_system_discrete} at each stage $i \Delta t$.

\begin{figure}[b]
	\centering
	\includegraphics[width=1\linewidth]{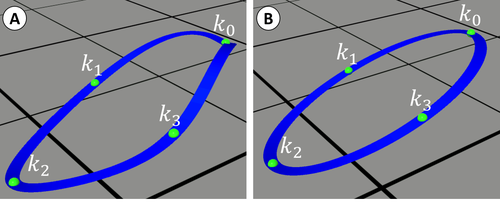}
	\caption{Same trajectory, optimized to only follow the keyframes (A) and to follow keyframes
		as well as minimizing snap on each stage of the optimization problem (B).}
	\label{fig:position_vs_derivatives}
\end{figure}

The first goal of our optimization scheme is then to follow the user inputs as closely as possible,
expressed by the cost
\begin{equation}
\label{eq:keyframe_cost}
E^{k} = \sum_{j=1}^{M} || r_{\eta(j)} - k_{j} || ^ {2} .
\end{equation}
A small residual of $E^{k}$ indicates a good match of the planned quadrotor position and the specified
keyframe. The bounds \eqref{eq:input_bounds} together with \eqref{eq:quad_dynamical_system_discrete} and \eqref{eq:keyframe_cost} can then be formulated as a quadratic program
\begin{align}
	\label{eq:quadratic_program}
	\underset{X}{\text{minimize}} \ & \frac{1}{2} X^{T} H X + f^{T} X  \\
	\text{ subject to } & A_{\mathit{ineq}} X \leq b_{\mathit{ineq}} \nonumber \\
	\text{ and } & A_{\mathit{eq}} X = b_{\mathit{eq}} \nonumber ,
\end{align}
where $X$ denotes the stacked state vectors $x_{i}$ and inputs $u_{i}$ for each time-point, $H$ and $f$ contain the
quadratic and linear cost coefficients respectively which are defined by \eqref{eq:keyframe_cost}
, $A_{\mathit{ineq}}$, $b_{\mathit{ineq}}$ comprise the linear inequality
constraints of the inputs \eqref{eq:input_bounds} and
$A_{\mathit{eq}}$, $b_{\mathit{eq}}$ are the linear equality constraints
from our model \eqref{eq:quad_dynamical_system_discrete}
for each time-point $i \in {1,\ldots,N}$.
This problem has a sparse structure and can be readily solved by most optimization software
packages. However, this problem
is ill-posed and the result for a particular set of keyframes might be counterintuitive at first.
Since we only measure the match of quadrotor position at the keyframe times the state at other
time-points is not constrained in any way except for the quadrotor dynamics. Therefore a straight
path between two keyframe positions is as good as a zig-zag pattern if it is feasible.
An example of this is shown in \figref{fig:position_vs_derivatives}.
To attain better results we have to further regularize the solution.

In many robotics application one goal is to minimize energy expenditure and this is often done by
penalizing non-zero inputs or in other words attempting to reach desired positions with minimal
wasted effort. For end-user applications, for example in the context of a racing game, one can also
aim to attain smooth trajectories by penalizing higher derivatives of the quadrotor's position with
respect to time such as acceleration (2nd) or jerk (3rd).
We introduce the cost
\begin{equation}
\label{eq:quad_derivative_cost}
E^{d} = \sum_{i=q}^{N} || D^{q} \begin{bmatrix}
x_{i} \\
\ldots \\
x_{i-q} \\
\end{bmatrix} || ^ {2} ,
\end{equation}
where $D^{q}$ is a finite-difference approximation of the $q$-th derivative from the last $q$ states.
Since the term jerk is not commonly known outside of engineering fields an intuition is to think of
high values of jerk as a feeling of discomfort caused by too sudden motion. Humans tend to plan
motion by minimizing the norm of jerk \cite{flash1985coordination} and thus, minimizing jerk results in motion plans
that appear pleasant to a human.

The combined cost $E = \lambda_{k} E^{k} + \lambda_{d} E^{d}$ with weights $\lambda_{k|d}$ is
still a quadratic program and enables us to generate trajectories that are feasible and
that are optimal in the sense of \eqref{eq:quadratic_program}.
While still relatively basic in functionality this already enables a variety of use-cases such as aerial
light-shows and racing-games as illustrated in the next section.

\subsection{Optimizing for Human Objectives}
With the basics in place we now turn our attention to including high-level human objectives into the
optimization. As a running example we will consider the task of planning an aerial video-shot but we
would like to emphasize that many other tasks such as 3D reconstruction or projector based
augmented reality could be implemented in the same way.

We have already discussed how this process works from the user's perspective in the system overview.
Here the user provides additional camera targets that should be recorded at a specific time
(see \figref{fig:sys_overview_camera}). Furthermore, we assume that the quadrotor is equipped with a
gimbal that we can control programmatically. From a cinematographic standpoint, the most pleasant
viewing experience is conveyed by the use of either static cameras, panning ones mounted on tripods
or cameras placed onto a dolly (cf.~\cite{mascelli1998five}). Changes between these shot types can
be obtained by the introduction of a cut or jerk-free transitions, i.e. avoiding sudden changes in
acceleration. Furthermore, it is desirable to introduce saliency constraints or in other words we
want not only the camera path to be smooth but also want to keep the target motion within the image
frame as steady as possible and constrain it's motion to smooth motion.

\begin{figure}
	\centering
	\includegraphics[width=1.0\linewidth]{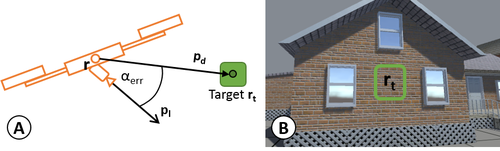}
	\caption{(A) camera direction $\mathbf{p_{l}}$ and distance $\mathbf{p_{d}}$. (B) effect of minimizing camera angle error $\alpha_{err}$ w.r.t. the target $\mathbf{r_{t}}$
		in the center of the FOV of the camera.}
	\label{fig:camera_angle_error}
\end{figure}

To achieve these high-level objectives, we include a target position for each stage into the
optimization variable. Analogous to the quadrotor position we introduce a cost term $E^{t}$ that
measures the deviations of user-specified keytarget points from the target positions at the
corresponding stages. We penalize higher temporal derivatives (acceleration and jerk) of the target
position by including finite differences in the cost term $E^{t,d}$.
To link the quadrotor and target trajectories we introduce a simple gimbal model:
\begin{gather*}
\dot{\psi_{\mathit{g}}} = u_{\mathit{g},\psi} \\
\dot{\phi_{\mathit{g}}} = u_{\mathit{g},\phi} \nonumber \\
[\psi_{\mathit{g},\mathit{min}}, \phi_{\mathit{g},\mathit{min}}]^{T}
\leq [\psi_{\mathit{g}}, \phi_{\mathit{g}}]^{T}
\leq [\psi_{\mathit{g},\mathit{max}}, \phi_{\mathit{g},\mathit{max}}]^{T} \\
\mathbf{u}_{\mathit{g},\mathit{min}}
\leq [u_{\mathit{g},\psi}, u_{\mathit{g},\phi}]^{T}
\leq \mathbf{u}_{\mathit{g},\mathit{max}} ,
\end{gather*}
where the inputs $u_{\textit{g},\psi}$, $u_{\textit{g},\phi}$ represent the angular velocities
of the yaw $\psi_{\mathit{g}}$ and pitch $\phi_{\mathit{g}}$ of the gimbal
and both the inputs and the absolute angles are bounded according to the physical gimbal.
The bounds specify the limits on the absolute angles and the angular velocities.
To ensure a smooth motion of the gimbal we introduce a cost
$E^{\mathit{g}}$ on temporal finite differences of the yaw and pitch angles analogous to
\eqref{eq:quad_derivative_cost}.
We do not incorporate the attitude of the quadrotor into our gimbal model and therefore the
bounds have to be chosen conservatively.

The angle between the current camera direction $\mathbf{p_{l}}$ and the direction of the target
$\mathbf{p_{d}}$ is depicted in \figref{fig:camera_angle_error}. The error is
then computed by
\begin{gather}
	\label{eq:camera_angle_error}
	\alpha_{\mathit{err}} = \cos^{-1} \left( \frac{\mathbf{p_{d}} \cdot \mathbf{p_{l}}}{| \mathbf{p_{d}} | | \mathbf{p_{l}} |} \right) \\
	\mathbf{p_{d}} = \mathbf{r_{t}} - \mathbf{r} \text{ and } 
	\mathbf{p_{l}} = \begin{bmatrix}
	\cos \phi_{g} \cos(\psi + \psi_{g}) \\
	\cos \phi_{g} \sin(\psi + \psi_{g}) \\
	\sin \phi_{g} \\
	\end{bmatrix} \nonumber ,
\end{gather}
where $\mathbf{r}$ is the quadrotor position, $\mathbf{r_{t}}$ is the target position
and $\psi$ is the pitch angle of the quadrotor.

Deviations of the camera direction from the desired target are penalized by
\begin{equation}
E^{c} = \sum_{i=1}^{N} \left( \alpha_{\mathit{err}}^{i} \right)^2 ,
\end{equation}
where $\alpha_{\mathit{err}^{i}}$ is the camera angle error at stage $i$.
Here the separation of target trajectory from the camera direction might seem surprising but it gives more
flexibility as the user can choose the weights of the importance of target keypoints and the camera direction separately.

\begin{figure}[tb]
	\centering
	\includegraphics[width=1.0\linewidth]{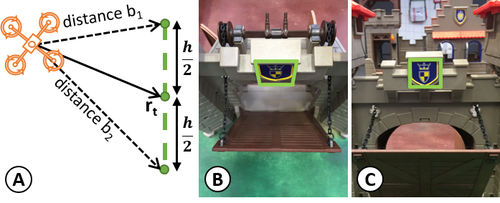}
	\caption{(A) Illustration of skewness error, where $b_{1}$ and $b_{2}$ are the distances to the upper
		and lower edge of the target bounding box. (B+C) Perspective without skewness correction (B) and
		with (C). Note that target is centered in both images.}
	\label{fig:skewness_error}
	\vspace{-3pt}
\end{figure}

The final aesthetic cost is related to perspective effects. Viewpoints that are to high or low relative
to the recorded object of interest lead to skew and results in strong vanishing lines in the image. This is
illustrated in \figref{fig:skewness_error}. While this effect maybe desired in some situations
(imagine an overhead shot) we allow the user to supress these types of distortions by optionally including a skewness cost $E^{s}$: 
\begin{gather}
	s_{\mathit{err}} = \frac{b_{1}}{b_{2}} - 1 = \frac{
		(\mathbf{p_{d}} + \mathbf{\tilde{h}/2}) \cdot \mathbf{p_{d}}
	}{
	(\mathbf{p_{d}} - \mathbf{\tilde{h}/2}) \cdot \mathbf{p_{d}}} - 1 \nonumber \\
\mathbf{\tilde{h}} =
\begin{cases}
	\mathbf{h} , & \text{ if } p_{d,3} >= p_{t,3} \\
	\mathbf{-h} , & \text{ else }
\end{cases} \nonumber , \\
E^{s} = \sum_{i=1}^{N} \left( s_{\mathit{err}}^{i} \right)^2 \\
\label{eq:skewness}
\end{gather}
where $\mathbf{h}$ is a vertical vector pointing from the center of the target to the upper edge of the bounding box
and $s_{\mathit{err}}^{i}$ is the skewness error at stage $i$. In the computation we distinguish the case of a quadrotor flying above the target and the
case of flying below a target.

Summing up the individual cost terms gives results in the final cost
$E = \sum_{i=\left\{k, s, t, g, c\right\}} \lambda_{i} E^{i}$
Unfortunately $\alpha_{\mathit{err}}$ and $s_{\mathit{err}}$ are non-linear in the variables of the motion
plan and in consequence minimizing $E$ can no longer be written as a quadratic program. We describe how we
minimize $E$ in the implementation section.

By penalizing snap of the quadrotor position and jerk of the camera motion the combined cost results in
aesthetically pleasing footage (see the accompanying video). We can now generate a motion plan for a
quadrotor that follows a target trajectory with the camera.
To further support novice users we included an approximate collision-free scheme that
can be used to keep a minimum distance from the target or stay at a safe distance from
obstacles. Again we refer to the implementation section for details.
Note that this only works for static objects where the position is known at the time of trajectory generation.

\section{IMPLEMENTATION}
In this section we describe how we implemented the different components of our
system. We start with describing the iterative quadratic programming scheme, then
explain the onboard controller and the quadrotor hardware and finally show how we
realized the design tool. 

\begin{figure*}[t]
	\centering
	\includegraphics[width=1.0\linewidth]{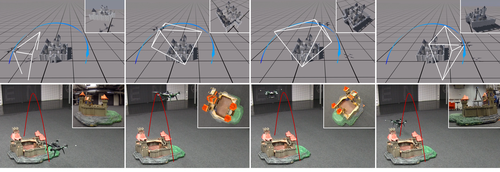}
	\caption{Aerial camera shot of a toy castle. Top row: planned trajectory in our design tool. Bottom row: flown trajectory.}
	\label{fig:apps_cinematography_dolly}
\end{figure*}

\paragraph{Iterative Quadratic Programming} To solve the non-linear problem described
above we resort to a scheme of iterative quadratic programming (IQP).
The general idea is to linearly or quadratically approximate the problem around the
current estimate of the
solution. This approximate system is then solved and a better, consistent estimate of
the solution is found.
These iterative schemes usually converge within a few iterations despite the cost
functions not being convex
anymore. In our concrete implementation we start with an initial guess of the
trajectory by interpolating
the quadrotor positions and the camera targets between the keyframes. We also
enforce all initial equality
constraints to be fulfilled. As the proposed energies are usually non-convex a good
initial guess is important
to find a good solution. For each major iteration of our solver we build the $H$ and
$f$ matrices of a quadratic program. This is done by quadratically approximating each
of the cost terms around
the trajectory $X$, note that this does not affect the quadratic terms in $E$.
We also assemble the
bounds and equality and inequality constraints and linearize them analogously. The
fully assembled system is
a sparse quadratic program and can be solved by most optimization packages. The
solution gives us a change
$dX$ of the current motion plan. We perform a line search with the step length
$\alpha$ to find a new motion
plan $X_{\mathit{next}} = X + \alpha dX$. We lower $\alpha$ until we find a
$X_{\mathit{next}}$ with a cost
$C(X_{\mathit{next}}) < C(X)$. This step is necessary as the cost of the approximated
quadratic program is only an approximation of the real residual. An empirically derived serves as termination criteria.
 
\paragraph{Obstacle Avoidance:} We approximate each obstacle as a static sphere with a radius
$o_{r}$ and introduce a non-convex constraint $|| r \geq o_{r} || ^ {2}$.
We linearize these constraints in each IQP iteration for each
stage in time around the current trajectory $X$.
Although we cannot guarantee global optimality of the resulting trajectories, this approach
can be helpful for planning trajectories in scenes with known geometry and many objects. More advanced collision avoidance schemes, potentially taking dynamic targets into consideration (e.g., \cite{Javier2015Collision}), could be included in future work. 

\paragraph{Algorithm Performance:} To evaluate the performance of our optimization scheme we measured the time necessary to generate different trajectories. The runtime of the algorithm depends on the flight duration, the number of keyframes and the constraints which are incorporated into the optimization problem. Typical run times (Intel Core i7 4GHz CPU, Matlab's quadprog solver) are $\sim$1 sec for pure QPs (e.g., the trajectory in \figref{fig:apps_light_writing} had a flight time of 30 sec and was generated in 1.4 sec) and tens of sec for IQPs (e.g., the trajectory in \figref{fig:apps_cinematography_dolly} had a flight time of 20 sec and was generated in 14 sec). Optimizing over a receding horizon which is shifted along the trajectory may be a fruitful strategy to speed up the algorithm. Another idea would be to split a trajectory in overlapping and reasonable constrained sub-trajectories and optimize them separately. Both approaches would negate the global optimality property of generated trajectories, requiring evaluation of real-world feasibility.

\paragraph{Onboard Control and Hardware:} Once we generate trajectory control inputs
these can be transmitted to a real quadrotor.
Our real-time control system builds on the PIXHAWK autopilot
software \cite{Meier2012}.
Desired positions along the motion plan, camera look-at vectors and target
trajectories are transmitted from a ground-station via the Robot Operating System
(ROS). An LQR (Linear-quadratic regulator) running on a
dedicated single-board computer computes the necessary forces and moments to track
the motion plan\hidden{as well as compensates for air disturbances and other sources of error}.
These forces are then translated into low-level
rotor and gimbal speeds by further controllers running on a PX4 FMU. We created result figures using two different quadrotor platforms: the 3DR Solo and a custom-build Pixhawk-based platform.

\textit{Design Tool:} The 3D trajectory design tool has been implemented as Unity 3D
tool which allows for easy adaptation and
integration of a variety of IO devices. A further advantage of this design decision is
that it is easy to develop
augmented reality applications such as mixing real and virtual quadrotors in a
racing scenario. We have
interfaced the design tool with our optimization algorithm implemented using the
Matlab optimization toolbox. 
The source code for the optimization algorithm can be found in the supplementary materials as
self-contained Matlab code.

\section{RESULTS AND APPLICATION SCENARIOS}\label{sec:results}
Despite having used camera planning as running example we note that our method is general and can be applied in many different application scenarios. In particular, the discrete nature of the proposed IQP scheme makes it straightforward to incorporate application specific constraints. In this section we want to illustrate a number of interactive usage scenarios which we have implemented using our method.  

\subsection{Light Painting}
Quadrotors have already been used in entertainment settings, in particular to create spectacular aerial light shows (cf. \cite{ArsElectronica2012}). However, creating such complex and coordinated flight patterns is not possible with consumer grade technologies and hence has not been accessible to the end-user. Our tool allows for straightforward end-user design of such creative scenarios. 

\begin{figure}[tb]
	\centering
	\includegraphics[width=1.0\linewidth]{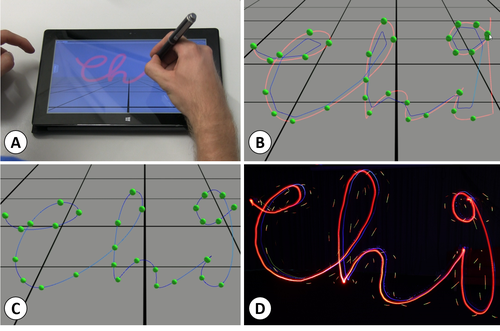}
	\caption{(A) Handwritten input. (B) Initial feasible trajectory can be overly smooth. (C) After iteration a feasible and visually pleasing trajecory is found. (D) Final result flown by MAV and captured via long-exposure photography.}
	\label{fig:apps_light_writing}
\end{figure}

One such example is illustrated in \figref{fig:apps_light_writing}. Here the user provides input position constraints by writing or sketching the desired shape. Our method then generates a feasible trajectory which as a side-effect of minimizing snap also smooths the input strokes.  However, the generated trajectory may not coincide with the desired output e.g. because it linearly interpolates the keyframes so that handwriting may not be legible anymore (see \figref{fig:apps_light_writing}, B). The user can correct for this by changing the parameters of the optimization scheme (e.g., weights of the energies) or by adjusting keyframe positions and timings. 

Once satisfied the trajectory can be flown by a real robot. In \figref{fig:apps_light_writing} we have mounted a bright LED to the robot and captured the flight path via long-exposure photography. %

\subsection{Racing}
Another interesting application domain is that of aerial racing. First person drone racing is an emerging sport that requires a lot of expertise in manual quadrotor control. Our tool can bring this within reach of the end-user. As a proof of concept we have developed a simple aerial racing game. In this scenario a user can design a free-form race course, specifying length, curvature and other parameters as well as overall race-time. 

\begin{figure}[b]
	\centering
	\includegraphics[width=1.0\linewidth]{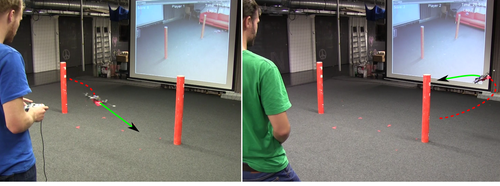}
	\caption{Two player aerial racing. User input is weighted with automatic control to adjust difficulty.}
		\label{fig:apps_racing}
\end{figure}

For the race itself we implemented a semi-manual flight mode for which we changed the position controller, by remapping the feedforward term ($m\mathbf{\ddot{r}}_d$) of \eqref{eq:positioncontrol} to the joystick of a game controller. Thereby, the user can choose the direction and the strength of the feedforward-force allowing him to deviate with the quadrotor from the generated reference trajectory. Users can then, for instance, take a short cut in a curve or fly the trajectory with a higher velocity than generated by the optimization method. The score is calculated as a function of the deviation from the generated trajectory and the time needed to complete all laps. In other words, the player who managed to stay on the trajectory as fast as possible will win. We note that by manipulating the underlying controller, it would be possible to introduce further video game concepts such as player strength balancing into real-world quad racing. 
For example, allowing a player to temporarily race on a faster reference trajectory than his opponents.

\subsection{Aerial Videography}
Our main results stem from the application scenario of aerial videography. We have already mentioned the technical details and how we incorporate cinematographic goals into our optimization scheme. Here we briefly summarize a number of interesting and challenging video-shots (best viewed in video).%

\figref{fig:apps_cinematography_dolly} illustrates a shot where a quadrotor flies over a toy castle and at the same time records it. Here the gimbal has to smoothly track the target just as the quadrotor swoops over the object and turns around its own axis once reaching the highest point. Such a shot composition is difficult to achieve manually due to the complicated quadrotor-camera-target coordination.

\begin{figure*}[t]
	\centering
	\includegraphics[width=1.0\linewidth]{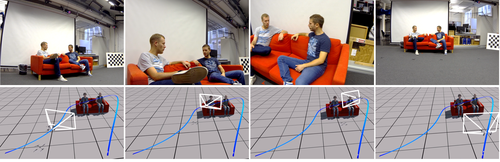}
	\caption{Multi target shot. Top row: frames of the video sequence shot by the onboard camera. Bottom row: according quadrotor positions shown in the preview of the design tool.}
	\label{fig:apps_cinematography_couch_slider}
\end{figure*} 

Even with conventional cameras, composition of multi target shots is a very challenging task. Aerial-videography makes this even more difficult due to the many
degrees of freedom and complex geometric dependencies requiring coordination for smooth, jerk-free transitions from one to the next target while airborne. 
In \figref{fig:apps_cinematography_couch_slider} we illustrate a sliding shot, transitioning between targets -- the two actors -- while the camera is moving from left to right and steadily rising in altitude. Throughout the entire trajectory the oerientations of quad and camera never remain constant, yet the camera targets are kept in focus and the transitions are smooth. Flying such a trajectory manually would only be possible with two operators, one for steering the camera, the other the quadrotor.

\section{DISCUSSION AND FUTURE WORK}\label{sec:discussion}
So far we presented a novel method to generate quadrotor trajectories subject to high-level goals and demonstrated its feasibility in different applications. In the remainder of this paper we are going to discuss the limitations of our approach and highlight interesting areas for future work.

\subsection{Limitations}
The optimization framework proposed in this paper has proven to be powerful and versatile however there are of course a number of limitations. First, our goal is to enable non-expert users to design arbitrary MAV use cases. While the method is generic and designed to be extensible it does require expertise and effort to formalize further objectives (that we have not treated so far) and to integrate them into the algorithm. We believe that our high-level design tool bridges the gap between the underlying optimization algorithm and end-user goals sufficiently well. Nonetheless it is an interesting future research question how end-users could extend not only the use-cases we have demonstrated but also the optimization itself.   

Currently all our application scenarios depend on a high precision indoor tracking system. This is a limiting factor as one would of course like to fly many of the examples outdoors using GPS sensing. To this end our method is generic and could be made to work with any localization system, in particular with GPS position data in outdoor scenarios. However, we have not implemented this and of course the localization accuracy would impact the exact results. 

\subsection{Future Work}
The optimization-based design of quadrotor trajectories subject to high-level user constraints is a comprehensive research space and our work only started to cover it. The investigations we did so far raised a number of additional research problems. In our method, the time at which a certain keyframe is reached cannot be changed by the optimization scheme. To extend our algorithm, it would be interesting to formalize the optimization problem in a way that both, the keyframe's position and its time can be optimized.

Furthermore, the use case of aerial videography raised the question: what is an aesthetic aerial video sequence and is it possible to optimize for it? By incorporating human objectives into the optimization, our work already presents a starting point, nevertheless it would be interesting to see whether further concepts and rules of cinematography can be incorporated into the optimization problem. 

Nonetheless, our method is generic and can be applied to further use cases. For example, it would be possible to extend the method to 3D scanning of buildings and other objects of interest. Here one could integrate objectives that capture e.g., reconstruction quality and surface coverage. Another possible example includes an advanced flying action camera for outdoor usage enabling users to trigger pre-defined trajectories on-the-fly. These scenarios would obviously require additional information such as the environment's 3D geometry for collision avoidance and accurate localization of drone and human in the outdoor case. Finally, it is not only drones that our method applies to. Most straightforward would be an extension to other actuated camera platforms such as dollies and robotic arms.

\section{CONCLUSION}
In summary we have proposed a user in the loop design tool for the creation of aerial robotic behavior. At its core lies an optimization-based algorithm that integrates low-level quadrotor control constraints and high-level human objectives. Therefore, we used a linear approximation of the quadrotor model enabling us to generate trajectories subject to the physical limits of a quadrotor. Stating the problem as discrete, additionally permits the easy incorporation of high-level constraints to support the user, for instance, in the creation of pleasing aerial footage. This allows users to concentrate on the creative and aesthetic aspects of the task at hand and requires little to no expertise in quadrotor control or the target domain. We have demonstrated the flexibility and utility of our approach in three different use cases including aerial videography, light painting and racing.

\section{ACKNOWLEDGEMENTS}
We thank C\'ecile Edwards-Rietmann for providing the video voiceover and G\'abor S\"or\"os for creating the aerial light-painting videos.
We are also grateful for the valuable feedback from the associate chairs and external reviewers. This work was partially funded by the Swiss Science Foundations (UFO 200021L\_153644) and Microsoft Research.

\bibliographystyle{SIGCHI-Reference-Format}
\bibliography{bib/mav,bib/ufo,bib/dc,bib/video_stab,bib/misc}

\balance

\begin{appendix}
\label{sec:appendix}
In the work proposed here we use an approximation of a full, non-linear quadcopter model for the optimization-based generation of  trajectories. However, the resulting trajectories need to be flown by a real quadcopter and hence one must relate the approximate model to the full model of the quadcopter. Here we briefly summarize the modeling and control aspects necessary for replication of our method. A full introduction to this topic is beyond the scope of this work and we refer the interested reader to \cite{Mahony2012}.

\subsection{Quadrotor Model}
A quadrotor is a robot with four identical rotors which generate a thrust and a moment orthogonal to
the square they span. Our quadrotor model closely follows \cite{Mellinger2011a}.
To describe the configuration of a quadrotor we define its position as the
location of the center of mass in an inertial world coordinate frame
($\mathbf{x}_{W}$, $\mathbf{y}_{W}$, $\mathbf{z}_{W}$),
and its attitude as the rotation of the body-fixed frame
($\mathbf{x}_{B}$, $\mathbf{y}_{B}$, $\mathbf{z}_{B}$) with
respect to the world frame (see \figref{fig:quadrotor}). The rotation matrix from body
to world frame is then given by
$R_{BW} = [\mathbf{x}_{B}\ \mathbf{y}_{B}\ \mathbf{z}_{B}] \in \mathbb{SO}(3)$.
Each rotor of the drone has an angular speed
$\omega_i$ and produces a force $F_{i}$ and moment $M_{i}$, according to
\begin{equation*}
F_i = k_{F} \omega_{i}^2 , \quad M_{i} = k_M \omega_{i}^2 ,
\end{equation*}
where $k_F$ and $k_M$ are constants specific to the rotors. Therefore, the control input to the
quadrotor can be written as $\mathbf{u}$ where $u_1$ is the net force in
$\mathbf{z}_{B}$ direction and
$u_2$, $u_3$, $u_4$ are the moments in $\mathbf{x}_{B}$, $\mathbf{y}_{B}$, $\mathbf{z}_{B}$ direction
acting on the quadrotor.
The input can be expressed in terms of
the rotor speeds $\omega_1$, $\omega_2$, $\omega_3$, $\omega_4$:
\begin{equation}
\begin{bmatrix}
 u_1 \\
 u_2 \\
 u_3 \\
 u_4
\end{bmatrix}
=
\begin{bmatrix}
 k_F & k_F & k_F & k_F \\
 0 & k_FL & 0 & -k_FL \\
 -k_FL & 0 & k_FL & 0\\
 k_M & -k_M & k_M & -k_M
\end{bmatrix}
\begin{bmatrix}
 \omega_1^2 \\
 \omega_2^2 \\
 \omega_3^2\\
 \omega_4^2
\end{bmatrix} ,
\label{eq:quadrotorinputs}
\end{equation}
where L is the distance from the axis of rotation of the rotors to the center of mass of the quadrotor.

The position of the quadrotor in the world frame can be specified according to Newton's
equation of motion governing the acceleration of a mass point:
\begin{equation}
m\ddot{\mathbf{r}} = u_1 \mathbf{z}_{B} + m \mathbf{g} \in \mathbb{R}^3 ,
\label{eq:equation_of_motion_masspoint}
\end{equation}
where $\mathbf{r}$ is the position vector, $\mathbf{g} = [0, 0, -g]^{T}$ is the gravity vector pointing along the $-z$ axis of the
world frame, $g$ is the gravitational constant and $m$ is the mass of the quadrotor.

\begin{figure}[b]
	\centering
	\includegraphics[width=0.9\linewidth]{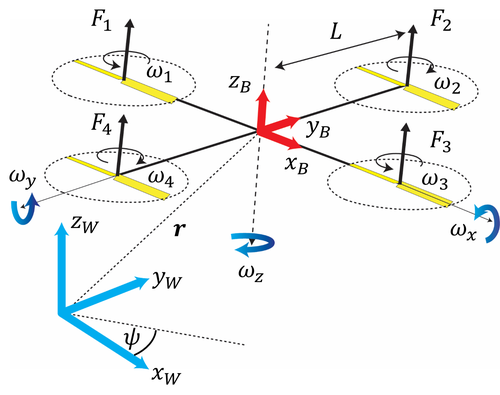}
	\caption{A quadrotor in 3D with its flat outputs (position $\mathbf{r}$, yaw angle $\psi$), 
		world ($\mathbf{x}_{W}$, $\mathbf{y}_{W}$, $\mathbf{z}_{W}$) and body frame ($\mathbf{x}_{B}$, $\mathbf{y}_{B}$, $\mathbf{z}_{B}$), the rotational velocities of the quadrotor in each dimension ($\omega_x$, $\omega_y$, $\omega_z$), the distance $L$ from the axis of rotation of a rotor to the center of mass of the quadrotor, as well as the thrust forces $F_i$ and angular velocities $\omega_i$ of each rotor.}
	\label{fig:quadrotor}
\end{figure}

The Euler rotation equations are
\begin{equation}
\mathbf{M}_{B} = I \dot{\omega}_{BW} + \omega_{BW} \times I \omega_{BW} \in \mathbb{R}^3 ,
\label{eq:euler_rotation_equation}
\end{equation}
where $\mathbf{M}_{B} = [u_{2}, u_{3}, u_{4}]^{T} = [\omega_x, \omega_y, \omega_z]^{T}$ is the moment vector acting on the quadrotor in
the body frame, $\omega_{BW}$ is the angular velocity of the body frame in the world frame
and $I$ is the moment of inertia of the quadrotor in the body frame.

\subsection{Quadrotor Control}
As can be seen from the equations \eqref{eq:quadrotorinputs},
\eqref{eq:equation_of_motion_masspoint} and \eqref{eq:euler_rotation_equation},
the quadrotor configuration has $6$ degrees of freedom but only $4$ actuators.
Therefore it is an underactuated system and cannot follow arbitrary trajectories in
the configuration space.
However, Mellinger et al.\ show that the system is
\textit{flat} \cite{faiz2001trajectory} with respect to the
$4$ \textit{flat outputs}
$[\mathbf{r}, \psi]^{T}$ and thus a quadrotor can follow trajectories in this space, given that the corresponding inputs are bounded to values that the quadrotor can achieve \cite{Mellinger2011a}.
This flat output space is the configuration of our approximate quadrotor model.
 
We use the linear controller from \cite{Javier2015Collision} to generate the corresponding inputs for the quadrotor.
The desired thrust along the $z$-axis of the body frame is computed as
\begin{equation}
\mathbf{F}_{d} = -K(\mathbf{x} - \mathbf{x}_{d}) + m(g\mathbf{z}_w + \mathbf{\ddot{r}}_{d}) ,
\label{eq:positioncontrol}
\end{equation}
where $x=[\mathbf{r},\mathbf{\dot{r}}]^T$ is the actual and $\mathbf{x}_d$ the desired position and velocity of the quadrotor and $m(g\mathbf{z}_w +\mathbf{\ddot{r}}_d)$ the feedforward term which compensates for gravity and known accelerations. 
The state feedback matrix $K$ is computed using a linear quadratic control strategy with integral action. 

The desired force $F_{d}$ already defines two degrees of freedom of the quadrotor attitude. Using the nonlinear control strategy on $SO(3)$ described in \cite{lee2013nonlinear} we employ the desired yaw angle $\psi_d$ to compute the desired
attitude $R_{BWd}$ of the quadrotor:
\begin{align*}
	\mathbf{z}_{Bd} & = \frac{\mathbf{F}_{d}}{||\mathbf{F}_{d}||} \\
	\mathbf{y}_{Bd} & = \frac{\mathbf{z}_{Bd} \times [cos(\psi_{d}), sin(\psi_{d}), 0]^T}{|| \mathbf{z}_{Bd} \times [cos(\psi_{d}), sin(\psi_{d}), 0]^T||} \\
	\mathbf{x}_{Bd} & = \mathbf{y}_{Bd} \times \mathbf{z}_{Bd} \\
	R_{BWd} & = [\mathbf{x}_{Bd},\mathbf{y}_{Bd},\mathbf{z}_{Bd}] ,
\end{align*}
where
$\mathbf{y}_{Bd}$ are the desired $x$- and $y$-axis of the body frame.
To control the attitude we can now calculate the desired moment vector $\mathbf{M}_{Bd}$ in
$\mathbf{x}_{B},\mathbf{y}_{B},\mathbf{z}_{B}$ direction,
\begin{align*}
	\mathbf{e}_R & = \frac{1}{2} \mathit{vee} \left( R_{d}^{T} R_{BW} - R_{BW}^{T} R_{d} \right) \\
	\mathbf{e}_{\omega} & = R_{BW}^{-1} \left( \omega_{BW} - \omega_{BWd} \right) \\
	\mathbf{M}_{Bd} & = -K_R\mathbf{e}_R - K_{\omega}\mathbf{e}_{\omega} ,
\end{align*}
where $R_{BW}$ is the actual rotation of the quadrotor, $\mathbf{e}_R$ is the rotation error,
$\omega_{BW}$, $\omega_{BWd}$ are the angular and desired angular velocity, $\mathbf{e}_{\omega}$ is
the angular velocity error and $\mathit{vee}$ is the \textit{vee map} from
$so(3) \rightarrow \mathbb{R}^{3}$.
From $\mathbf{F}_{d}$ and $\mathbf{M}_{Bd}$ we can calculate the input $u$
and thereby the velocities of the rotors needed
to reach the desired position and yaw angle:
\begin{align}
	\label{eq:u1calculation}
	u_1 &= \mathbf{F}_{d} \cdot \mathbf{z}_{B} \\
	[u_2,u_3,u_4]^T &= \mathbf{M}_{Bd} \nonumber ,
\end{align}
where \eqref{eq:u1calculation} is the projection of the desired thrust $\mathbf{F}_{d}$ on the actual
z-vector of the body frame $\mathbf{z}_{B}$. Finally, using \eqref{eq:quadrotorinputs}
we can compute the angular velocities $\omega_{i}$ corresponding to the input $u$.

\subsection{Validity of Approximate Quadrotor Model}
Following the approach in \cite{Naegeli2014IROS}, we assume that the rotational dynamics of a quadrotor are fast compared to its
translational dynamics thus we can describe the behavior of the quadrotor by the thrust
vector $\mathbf{u}_{r}$ and the moment $u_{\psi}$ along the body-frame z-axis.
In the following we will only refer to the norm $u_{r}$ of the force vector
$\mathbf{u}_{r}$

Let $F_{\max}$ be the maximum force and $M_{\max}$ be the maximum moment each motor can produce.
Then the bound on the maximum possible thrust that the quadrotor can
achieve (i.e.\ all motors full on) is
\begin{equation*}
u_{r} \leq u_{r,\max} = 4 F_{\max}
\end{equation*}
and the bound on the maximum possible moment (i.e.\ two motors rotating in same
direction full on and the other two off) is
\begin{equation*}
u_{\psi} \leq u_{\psi,\max} = 2 M_{\max} .
\end{equation*}
Because the force and moment are coupled it is not possible to
achieve full thrust $u_{r,\max}$ and full moment
$u_{\psi,\max}$ at the same time.

Let us now assume a stricter bound on the maximum moment of the quadrotor:
\begin{equation*}
u_{\psi} \leq u_{\psi,\mathit{lim}} = \beta u_{\psi,\max}
\label{eq:moment_bound}
\end{equation*}
where $\beta \in [0, 1]$.
If we want to be able to achieve a moment of $u_{\psi,\mathit{lim}}$ at
all times we have to take into account that in the extreme case
two motors will be limited to a force of
$F_{\mathit{lim}} = (1 - \beta) F_{\max}$
and thus the bound on the thrust of the quadrotor is
\begin{equation*}
u_{r} \leq u_{r,\mathit{lim}} = (2 + 2 (1 - \beta)) F_{\max}
= \left( 1 - \frac{\beta}{2} \right) u_{r,\max}.
\label{eq:force_vector_bound}
\end{equation*}
For example, if $\beta = 0.2$, i.e.\ bounding the moment
to $20\%$ of the quadrotors maximum moment the quadrotor can still achieve $90\%$ of 
its maximum thrust at all times.
These limits still allow the agility of a quadrotor to be sufficient for  many use cases.

\end{appendix}

\end{document}